# Kinetics and Intermediate Phases in Epitaxial Growth of $Fe_3O_4$ Films from Deposition and Thermal Reduction

Xiaozhe Zhang[1,2], Sen Yang[1], Zhimao Yang[1], Xiaoshan Xu[2]

[1]School of Science, MOE Key Laboratory for Non-Equilibrium Synthesis and Modulation of Condensed Matter, Xi'an Jiaotong University, Xi'an 710049, ShaanXi, China

[2]Department of Physics and Astronomy, Nebraska Center for Materials and Nanoscience, University of Nebraska-Lincoln, Lincoln, NE 68588, USA

* Correspondence to: Xiaoshan Xu (xiaoshan.xu@unl.edu), Zhimao Yang (zmyang@xjtu.edu.cn)

## Abstract

We have studied the growth of $Fe_3O_4$ (111) epitaxial films on $Al_2O_3$ (001) substrates using a pulsed laser deposition / thermal reduction cycle using an $\alpha$-$Fe_2O_3$ target. While direct deposition onto the $Al_2O_3$ (001) substrates results in an $\alpha$-$Fe_2O_3$ epilayer, deposition on the $Fe_3O_4$ (111) surface results in a $\gamma$-$Fe_2O_3$ epilayer. The kinetics of the transitions between $Fe_2O_3$ and $Fe_3O_4$ were studied by measuring the time constants of the transitions. The transition from $\alpha$-$Fe_2O_3$ to $Fe_3O_4$ via thermal reduction turns out to be very slow, due to the high activation energy. Despite the significant grain boundaries due to the mismatch between the unit cells of the film and the substrate, the $Fe_3O_4$ (111) films grown from deposition/thermal reduction show high crystallinity.

# Introduction

Magnetite ($Fe_3O_4$), as one of the phases of iron oxides, has been of wide usage in, e.g., pigments, magnetic recording, and catalysis, due to their useful optical, magnetic, and chemical properties and the low cost. Additional applications are constantly being explored in $Fe_3O_4$, particularly in molecular biology and spintronics, because of its bio-compatibility and special electronic structures respectively.[1–6]

In the thermodynamic standard conditions, $Fe_3O_4$ is a metastable phase of iron oxide, which has an inverse spinel structure (space group Fd-3m).[7] In $Fe_3O_4$, the cubic unit cell contains eight formula units that can be written as $(Fe^{3+})_8[Fe^{3+}Fe^{2+}]_8O_{32}$. The $Fe^{2+}$ and $Fe^{3+}$ cations are located in the interstitial sites of oxygen anions sub-lattice. One cation site, occupied only by $Fe^{3+}$ ions, is tetrahedrally coordinated to oxygen. The other site, occupied by equal numbers of $Fe^{2+}$ and $Fe^{3+}$ ions, is octahedrally coordinated to oxygen. Below $T_N \approx 860$ K, $Fe_3O_4$ is ferrimagnetic, in which the magnetic moments of the $Fe^{3+}$ in two different sites cancels each other while the moments of the $Fe^{2+}$ are aligned and form a spontaneous magnetization.[8] The resistivity of $Fe_3O_4$ decreases when temperature increase, with a rapid change at $T_V \approx 120$ K, noted as the Verwey transition, which is often considered as a signature of the $Fe_3O_4$ phase.[9–13]

Great efforts have been devoted to preparing epitaxial $Fe_3O_4$ thin films using pulsed laser depositions, with a variety of target materials to begin with, including metal Fe,[14] $Fe_3O_4$,[9,15–18] and $Fe_2O_3$.[9–12,19] The iron oxide phase and oxygen stoichiometry of the films are sensitive to the growth conditions, especially temperature and background gas pressure.[15] Therefore, thermodynamics and chemical reactions on the substrate are the key for the growth of $Fe_3O_4$ films, as also indicated by many studies on the transitions between iron oxide phases using surface characterizations.[20–30] The surface structures, including termination, reconstruction, and morphology, have been the focus of study.[20–23,27] On the other hand, the kinetics of the transitions (time scale as a function of conditions) have seldom been systematically carried out, although the important energetic information can be extracted from the kinetics and the time scale itself is a critical factor both for studying or applying these transitions.

In this work, we studied the growth of $Fe_3O_4$ thin films using thermal reduction from the deposited $Fe_2O_3$ layers on $Al_2O_3$ (001) substrates. The time-resolved transitions between different phases of iron oxide were measured, using the reflection high energy electron diffraction (RHEED), around the phase boundaries. In particular, the transition from an α-$Fe_2O_3$ (001) layer to an $Fe_3O_4$ (111) layer show a long temperature-dependent time constant that follows the Arrhenius law with an activation energy of $2.3 \pm 0.6$ eV; the activation energy does not change significantly with the $O_2$ pressure while the time constant decreases with increasing pressure. These studies on kinetics of transitions between iron oxide phases at the surface are important for advancing our understanding on the response of the iron oxide surfaces to oxidation and reduction environments, which is critical in the application of iron oxides in heterogeneous catalysis, spintronics, and biomedicine.

## Experimental

The film deposition was carried out on 0.2° miscut and one side polished sapphire $Al_2O_3$ (001) substrates using pulsed laser deposition (PLD), in $5\times10^{-3}$ Torr $O_2$, with 600 °C substrate temperature. The substrates were heated by an infrared laser using an absorber attached to the back of the sample mechanically. Both the temperature of the absorber and the temperature of the substrate were monitored using two separate pyrometers. The uncertainty of the temperature measurements was about 50 °C. Before deposition, the substrates were annealed in base pressure (lower than 1×10$^{-7}$ Torr) at 600 °C for 30 minutes. Each deposition corresponds to an epilayer of approximately 2.5 nm. The target used for the deposition is $Fe_2O_3$ pellet prepared from high purity $Fe_2O_3$ powder, sintered at 1400 °C for 24 hours. An excimer laser (KrF, λ = 248 nm) was used at a fluence of 1.8 J/cm$^2$ and a repetition rate of 2 Hz. The target to substrate distance was kept at 5 cm. Thermal reduction of the α-$Fe_2O_3$ epilayers were carried out by heating the sample to high temperature in low $O_2$ pressure after the deposition of the target material. The transition between the iron oxide phases were studied using a time-resolved RHEED in which the images were taken automatically every 15 seconds. X-ray diffractions (XRD) were measured (θ-2θ scan) using a Rigaku D/Max-B diffractometer, with a cobalt K-$\alpha$ source ($\lambda = 1.79$ Å). The rocking curves were measured using a Rigaku SmartLab with a copper K-$\alpha$ source ($\lambda = 1.54$ Å). Magneto optical Kerr effect (MOKE) on the iron oxide films was measured using a He-Ne laser (632.8 nm) and a photoelastic modulator, in a longitudinal geometry. Atomic force microscopy (AFM) was measured using a Bruker Dimension ICON at room temperature. Temperature dependent resistivity was measured in a Janis cryostat using a Van der Pauw geometry.

## Results and Discussions

### Structural analysis of the epitaxial layers

First, we examine the different iron oxide phases that appeared during the growth, including hematite (α-$Fe_2O_3$), maghematite (γ-$Fe_2O_3$), and magnetite ($Fe_3O_4$). Figure. 2 shows the RHEED patterns of the iron oxide layers as well as the $Al_2O_3$ (001) substrate, where Fig. 2(a), (c), (e), and (g) correspond to the condition in which the electron beam points along the $Al_2O_3$ <001> direction, while Fig. 2(b), (d), (f), and (h) correspond to the condition in which the electron beam points along the $Al_2O_3$ <120> direction; the two directions are perpendicular to each other. These phases, as well as their epitaxial relations with the $Al_2O_3$ (001) substrate, are identified, according to the RHEED pattern, condition of appearance, and ex-situ characterizations. The structure of these phases and the epitaxial relations are depicted using models in Fig. 3 and summarized in Table I.[7,31–33] Below, we show more detailed analysis.

Direct deposition of the target material onto the $Al_2O_3$ (001) surface resulted in the RHEED patterns displayed in Fig. 2(c) and (d), which are similar to those of the $Al_2O_3$ (001) surface, indicating a similar in-plane lattice structure. Using the pattern separation of the $Al_2O_3$ (001) as the calibration, one can calculate the in-plane lattice constant as $5.07 \pm 0.1$ Å, which matches the lattice constant of the α-$Fe_2O_3$ (001) surface (see Table 1) within the experimental error. Since $Al_2O_3$ and $Fe_2O_3$ are isomorphic (R-3c corundum structure), it is understandable that the most stable iron oxide epilayer on $Al_2O_3$ (001) without thermal reduction, is α-$Fe_2O_3$ (001). Therefore, we assign the structural phase that shows the RHEED patterns in Fig. 2(c) and (d) as α-$Fe_2O_3$ (001). The epitaxial relation is $Al_2O_3$ (001) // α-$Fe_2O_3$ (001) and $Al_2O_3$ [100] // α-$Fe_2O_3$ [100]), as shown in Fig. 3(a). The corresponding reciprocal indices are marked accordingly.

The $Fe_3O_4$ (111) layer were obtained by thermally reducing the deposited $Fe_2O_3$ epilayer at high temperature. After the $Fe_2O_3$ epilayer underwent thermal reduction, we observed the RHEED patterns in Fig. 2(g) and (h), which also indicate a surface of triangular lattice. Assuming that a $Fe_3O_4$ (111) epilayer is on top of the $Al_2O_3$ (001) substrates and that the RHEED patterns observed in Fig. 2(g) and (h) are from the bulk reciprocal space projected onto the (111) surface, one can calculate the in-plane lattice constant as $5.93 \pm 0.05$ Å. For a normal $Fe_3O_4$ (111) surface, the in-plane lattice constant is 5.924 Å, which is close to the observed value. Therefore, we assign the structural phase that shows these patterns as $Fe_3O_4$ (111). The epitaxial relation is $Al_2O_3$ (001)//$Fe_3O_4$ (111) and $Al_2O_3$ [100]//$Fe_3O_4$ [-211], as shown in Fig. 2. There is no obvious match between the in-plane lattice constant of $Al_2O_3$ (001) and $Fe_3O_4$ (111), since the difference is more than 10%. To understand this epitaxial relation, we projected the $Al_2O_3$ unit cell and the $Fe_3O_4$ onto the (001) and (111) planes respectively, and overlapped the two unit cells, as shown in Fig. 3(b). The oxygen network appears to be overlapping well, which may be the reason for the epitaxial relation. The reciprocal indices of $Fe_3O_4$ (111) layer are marked in Fig. 2(g) and (h). Note that $Fe_3O_4$ has a face centered cubic (fcc) structure, so the lattice constant of the primitive cell of the (111) epilayer is $\frac{1}{\sqrt{2}}$ of the cubic lattice. Because of the fcc structure of the $Fe_3O_4$, only the reciprocal indices that have all-odd or all-even indices using the cubic indices are present, as shown in Fig. 2(g) and (h). Further verification of the $Fe_3O_4$ phase is found from ex-situ characterizations, such as x-ray diffraction, electric transport measurements, and magneto optical Kerr effect measurements (see the Subsection **Characterization of the $Fe_3O_4$ and $\gamma$-$Fe_2O_3$ films**).

The $\gamma$-$Fe_2O_3$ (111) epilayers were observed after deposition of the target material onto the $Fe_3O_4$ (111) surface. Among the iron oxide structural phases, $\gamma$-$Fe_2O_3$, another metastable phase of iron oxide at the thermodynamic standard conditions, has a similar structure with $Fe_3O_4$, as shown in Fig. 3(c). This structure can be represented as: $(Fe^{3+})_8[Fe^{3+}_{40/3}\square_{8/3}]O_{32}$ where $\square$ denotes vacancy, in which eight $Fe^{3+}$ atoms occupy tetrahedral sites while the remainder occupies octahedral sites.[34] In other words, it is a cation deficient spinel structure (space group $P4_332$). Figure. 2(g) and (h) shows the RHEED pattern of the epilayers after the deposition of target material onto the $Fe_3O_4$ (111) surface. Interestingly, these patterns differ dramatically from those in Fig. 2(e) and (f), suggesting that the surface structure has a determinant effect on the structure of the epilayer. Since the target is $Fe_2O_3$ which contains only $Fe^{3+}$, we expect mostly $Fe^{3+}$ in the epilayer after the direct deposition. Based on the structural similarity and valence consideration, a $\gamma$-$Fe_2O_3$ (111) epilayer is expected after the direct deposition of the target material onto the $Fe_3O_4$ (111) surface.[15,35] This is confirmed using a combined characterizations: in-situ RHEED and ex-situ x-ray diffractions (see Subsection **Characterization of the $Fe_3O_4$ and $\gamma$-$Fe_2O_3$ films**).

Having identified the $\gamma$-$Fe_2O_3$ phase, we also note that the RHEED patterns of the $\gamma$-$Fe_2O_3$ (111) epilayer are not consistent with the bulk reciprocal space projected onto the (111) plane. The structure of $\gamma$-$Fe_2O_3$ has similar lattice constants with those of $Fe_3O_4$. But because the lattice is simple cubic, the primitive cell is actually smaller; one expect no systematic extinction for the diffraction. This means that the RHEED patterns of the $\gamma$-$Fe_2O_3$ (111) epilayer are supposed to have more streaks than those of the $Fe_3O_4$ (111) epilayers. In contrast, the observed RHEED patterns of the $\gamma$-$Fe_2O_3$ (111) in fact show less streaks. As shown Fig. in 2(e), in the <01-1> direction (Fig. 2(f)), there appears to be no (022), (0-2-2) streaks for the $\gamma$-$Fe_2O_3$ (111) surface. In addition, along the <-211> direction, the (02-2) and (0-22) streaks are much weaker than the other diffraction streaks (marked as red arrow). In fact the RHEED patterns of the $\gamma$-$Fe_2O_3$ (111) epilayer is more consistent with a FeO (111) surface, which suggests a significant reconstruction at the $\gamma$-$Fe_2O_3$ (111) surface.[15]

The dependence of the structural phase on the structure of the beginning surface can be understood in terms of the interfacial energy. Since $\alpha$-$Fe_2O_3$ and $Al_2O_3$ are isomorphic, the energy of the $\alpha$-$Fe_2O_3$/$Al_2O_3$ interface is expected to be relatively lower than that of the $\gamma$-$Fe_2O_3$/$Al_2O_3$ interface. On the other hand, since $\gamma$-$Fe_2O_3$ and $Fe_3O_4$ have similar structures, the $\gamma$-$Fe_2O_3$/$Fe_3O_4$ interface is expected to have lower energy than that of the $\alpha$-$Fe_2O_3$/$Fe_3O_4$ interface. Therefore, after the direct deposition, the $\alpha$-$Fe_2O_3$/$Al_2O_3$ interface and the $\gamma$-$Fe_2O_3$/$Fe_3O_4$ interface are formed. In fact, a continuous change of epilayer structure from $\gamma$-$Fe_2O_3$ to $Fe_3O_4$ have been observed previously by changing the growth conditions.[15,27,35]

## Thermodynamics and the kinetics of the structural transitions

In order to study the kinetics of the $Fe_2O_3 \rightarrow Fe_3O_4$ transition, we first examined the boundary conditions between the phases using thermodynamic analysis. For both α-$Fe_2O_3$ (001) and γ-$Fe_2O_3$ (111) epilayers, the conversion to a $Fe_3O_4$ (111) layer at high temperature, involves not only a change of crystal structure, but also a loss of oxygen, which can also be treated as thermal reduction. The condition for the $Fe_2O_3 \rightarrow Fe_3O_4$ transition, can be estimated according to change of Gibbs free energy ($\Delta_r G$) in the following reaction:

$$Fe_2O_3 \rightleftharpoons \frac{2}{3} Fe_3O_4 + \frac{1}{6} O_2.$$

The $\Delta_r G$ for this reaction at certain temperature ($T$) and pressure ($P$) can be calculated from the Gibbs free energy at standard condition ($\Delta G^0$) of $Fe_2O_3$ and $Fe_3O_4$, using the relation

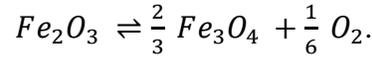

$\Delta_r G = \frac{2}{3}\Delta_f G^0_{Fe3O4} - \Delta_f G^0_{Fe2O3} + \frac{1}{6}RT \ln\left(\frac{P}{P_0}\right)$, (see the supplementary materials[36]),

where $P$ is the oxygen pressure and $T$ is the temperature. The standard formation Gibbs free energy ($\Delta_f G^0_{Fe2O3}$ and $\Delta_f G^0_{Fe3O4}$) can be calculated using the corresponding formation enthalpy ($\Delta_f H^0$) and formation entropy ($\Delta_f S^0$), which can be assume as constants. Table II and Table III[37] show the values of $\Delta_f H^0$ and $\Delta_f S^0$. The boundary between the α-$Fe_2O_3$ and the $Fe_3O_4$ phases is found by setting $\Delta_r G = 0$ and solving the relation between $P$ and $T$. As shown in Fig. 4(c), the solid line is the calculated phase boundary of the $Fe_2O_3$ and $Fe_3O_4$, consistent with the phase boundary calculated previously[29,38].

For the high pressure and low temperature region, α-$Fe_2O_3$ phase is stable, while for the low pressure and high temperature region, the $Fe_3O_4$ phase is stable. The above thermodynamic analysis provides the information on the boundary between the bulk α-$Fe_2O_3$ and $Fe_3O_4$ phases, but not the rate of the transition (kinetics). Various time scales have been mentioned during the studies on the transition between iron oxide phases;[20–30] however, a systematic study is lacking. Using the phase boundary in Fig. 4(c) as a guidance, we studied the kinetics of the α-$Fe_2O_3$→$Fe_3O_4$ transition for an epilayer on the $Al_2O_3$ (001) substrate, by measuring the time evolution of the structure during the thermal reduction using the time-resolved RHEED.

Starting from the $Al_2O_3$ (001) substrate, we deposited the target material (~2.5 nm), which generates an α-$Fe_2O_3$ (001) epilayer (see the supplementary materials[36]). During the thermal reduction of α-$Fe_2O_3$, we monitor the RHEED pattern with the incident electron along the <100> direction of the $Al_2O_3$. Images of the RHEED patterns were saved every 15 seconds and integrated along the longer dimension of the diffraction streaks (see the supplementary materials[36]). The evolution of the intensities of diffraction streaks is then plotted as a function of time. Figure. 4 (a) shows an example of RHEED intensity evolution at 930 °C in 9.2×$10^{-8}$ Torr $O_2$. We found that the $Fe_3O_4$ (02-2) diffraction streaks slowly emerged between the diffraction streaks of α-$Fe_2O_3$ (01) and (02), indicating the α-$Fe_2O_3$→$Fe_3O_4$ transition. The intensity of the $Fe_3O_4$ streaks increases and starts to saturate after a certain time (see Fig. 4(b)). We fit the intensity ($I$) of the $Fe_3O_4$ streaks using the formula $I = I_0(1 - e^{-t/\tau})$, where $t$ is time, $I_0$ is a saturation intensity, and $\tau$ is the time constant of the transition.

The time constant $\tau$ has been measured at different temperatures and the dependence is plotted in Fig. 4(d). When the temperature increases, $\tau$ decreases dramatically. We fit the temperature dependence of $\tau$ using the Arrhenius law, $\tau = \tau_0 e^{\frac{E_a}{kT}}$, where $E_a$ is the activation energy and $k$ is the Boltzmann constant. We repeated this study of time constant $\tau$ and activation energy at different $O_2$ pressure; the results are shown in Fig. 4(d). It is interesting that the activation energies do not change significantly considering the experimental uncertainty (see Table IV), while $\tau_0$ depends on the O$_2$ pressure dramatically.

In principle, the activation energy corresponds to the minimum energy barrier for the transition. For the Fe$_2$O$_3$→Fe$_3$O$_4$ transition, this energy barrier is related to breaking of the Fe-O bonds. The dissociation energy for a typical Fe-O bond is 4.2 eV however [39], which is about twice as much as the measured $E_a$ (2.3 ±0.6 eV on average). Therefore, it appears that at the surface, there are weaker Fe-O bonds that actually determine the $E_a$. In fact, the measured $E_a$ value is close to the band gap energy of α-Fe$_2$O$_3$.[40,41] The band gap energy of α-Fe$_2$O$_3$ corresponds to the energy to excite an electron from O back to Fe, which can be understood as the breaking of the weakest possible link between the Fe and O atoms.

The observation that the activation energy is not significantly affected by the pressure, is not surprising, since the change of O$_2$ pressure is not supposed to affect the Fe-O bond energy significantly. In principle, higher O$_2$ pressure means the condition is closer to the boundary between the Fe$_3$O$_4$ and α-Fe$_2$O$_3$ phases; intuitively, the time constant $\tau_0$ is expected to be larger. On the other hand, the experimental observations show that higher O$_2$ pressure actually makes the thermal reduction faster. We speculate a "fall-off" scenario of pressure dependent kinetics[42]: because the α-Fe$_2$O$_3$→Fe$_3$O$_4$ is endothermic, when the pressure is lower, the heat transfer is expected to be slower, which may affect the rate of the transition.

The Fe$_3$O$_4$→Fe$_2$O$_3$ transition is more complex. At high temperature, the time constant is much smaller for the transition. After the thermal reduction, we decreased the substrate temperature to 600 °C and increased the background O$_2$ pressure to 5×10$^{-3}$ Torr. The Fe$_3$O$_4$→α-Fe$_2$O$_3$ transition occurred within seconds, indicating a much smaller activation energy. Therefore, the rate of oxidation is not determined by the energy scale to break the O-O bonds (5.2 eV),[43] which is not untypical for reactions on transition metal oxide surfaces.[44] On the other hand, at lower temperature, the transition is not only slow, but with different direct product ($\gamma$-Fe$_2$O$_3$). We carried out the annealing of the Fe$_3$O$_4$ films in one atmosphere O$_2$ at 250 °C for 4 hours. The RHEED pattern after that turned out to be similar to the ones in Fig. 4 (e) and (f) (see the supplementary materials[36]). Ex-situ x-ray diffraction and magneto optical Kerr effect measurements indicates that the structure phase is $\gamma$-Fe$_2$O$_3$. (see the Subsection **Characterization of the Fe$_3$O$_4$ and $\gamma$-Fe$_2$O$_3$ films**).

## Characterization of the Fe$_3$O$_4$ and $\gamma$-Fe$_2$O$_3$ films

To confirm the structural analysis of the epitaxial layer, we have characterized the Fe$_3$O$_4$ films (~30 nm) ex-situ using x-ray diffraction, atomic force microscopy, electric transport, and magneto-optical Kerr effect.

Figure. 5(a) shows the $\theta$-$2\theta$ scan of the Fe$_3$O$_4$ film grown by repeating the deposition/thermal reduction cycles. No impurity phases can be identified from the diffraction peaks. Figure. 5(b) is the close-up view of the (111) peaks, where the Laue oscillation is obvious, indicating flat surface of the film. Fitting the Laue oscillation with the consideration of background,[45] one can find the film thickness as $25 \pm 1$ nm. The inset shows the rocking curve of Fe$_3$O$_4$ (111) peak, for which the full-width-half-maximum (FWHM) is 0.14 degree, indicating a high crystallinity. One can estimate the size of the crystallites of the film using the peak width of the rocking curve and the $\theta$-$2\theta$ scans; the results show a size of $57 \pm 1$ nm along in-plane direction and $26 \pm 1$ nm along the out of plane direction (see the supplementary materials[36]).

Figure. 6(a) shows the surface morphology of a $Fe_3O_4$ film measured using the atomic force microscopy. The surface of the film consists of domains separated by grooves of 1-2 nm deep. Despite this feature, the root mean square roughness of this film is 0.3 nm (Fig. S1(b)), confirming the flat surface indicated by the Laue oscillation observed in x-ray diffraction. Within the domains, the atomic terraces are observed (Fig. 6(c)), indicating again the high crystallinity. One of the origin of these grooves may be the unavoidable grain boundaries of the $Fe_3O_4$ films, when the epilayer has a larger unit cell than that of the substrate and the film nucleation occurs at different positions.[46–48] For example, when a $Fe_3O_4$ film is deposited on a MgO substrate, similar mechanism generates anti-phase boundary because the lattice constant of $Fe_3O_4$ is about twice as much as that of MgO.[46–48] The appearance of anti-phase boundary causes the deviation of transport properties and magnetic properties of the thin films from those of the bulk materials: the Verwey transition in the electric transport becomes less obvious with a decreased transition temperature; the magnetic coercivity is enhanced.[46–48]

To verify the Verwey transition in the $Fe_3O_4$ films, temperature dependence of the electrical resistance ($R - T$) has been measured between 50 and 300 K, as shown in Fig. 7(a). The Verwey transition temperature around 120 K is visible but not as clear as that in bulk,[10,12,13,48] consistent with the significant domain boundaries. To highlight the Verwey transition, we have calculated effective activation energy ($E_a^{eff}$) using the relation $E_a^{eff} = \frac{d \ln R}{d(\frac{1}{kT})}$, where $R$ is the resistance, $k$ is the Boltzmann constant; the result is shown in Fig. 7(b). A clear anomaly is observed at 114 K, which is attributed to the Verwey transition.

Using the magneto-optical Kerr (MOKE) effect, we have measured the in-plane hysteretic behavior of the magnetization ($M - H$) of the $Fe_3O_4$ films at room temperature, as shown in Fig.

8. One can identify two features of the M-H loop here: a coercivity of ~270 Oe and a saturation field larger than 1000 Oe. Both of the coercivity and the saturation field are much larger than the bulk values, which may come from the effect of the significant grain boundaries.[46–48]

To verify the observation of the γ-$Fe_2O_3$ as an intermediate phase in the deposition, we annealed a $Fe_3O_4$ film (~30 nm) in one atmosphere $O_2$ at 250 °C for 4 hours.[49] The RHEED patterns of the annealed film turn from those in Fig. 2(g) and (h) to those in Fig. 2(e) and (f), indicating a phase transition (see the supplementary materials[36]). As shown in Fig. 5(a), the x-ray diffraction spectrum of the annealed film is similar to that of the $Fe_3O_4$ film, except that the angles are systematically larger. The lattice constants calculated from the x-ray diffraction is 8.338 Å for the $Fe_3O_4$ film and 8.222 Å for the annealed $Fe_3O_4$ film, in agreement with the difference between the bulk values of $Fe_3O_4$ and γ-$Fe_2O_3$. Therefore, it appears that the annealed $Fe_3O_4$ film, as well as the layer observed in Fig. 2(e) and (f), are in fact the γ-$Fe_2O_3$ phase.

The hysteretic of the magnetization ($M - H$) of the γ-$Fe_2O_3$ films is shown in the Fig. 8. One can see that the coercivity of the γ-$Fe_2O_3$ film is significantly less than that of the $Fe_3O_4$ film, which is in line with the results in the previous studies.[15]

## Conclusion

By studying the growth of $Fe_3O_4$ (111)/$Al_2O_3$ (001) films using pulsed laser deposition and thermal reduction and studying the kinetics of the transitions between the iron oxide phases, we have found the activation energy for the α-$Fe_2O_3$→$Fe_3O_4$ transition is 2.3 ± 0.6 eV, corresponding to the weakest Fe-O bond to break at the surface. While the α-$Fe_2O_3$→$Fe_3O_4$ transition is slow due to the high activation energy, the $Fe_3O_4$→$Fe_2O_3$ transition is in general much faster and more complex. At high temperature, the oxidation of $Fe_3O_4$ is quick and results directly in the α-$Fe_2O_3$ phase; at lower temperature, the oxidation of $Fe_3O_4$ is much slower and generates the intermediate $\gamma$-$Fe_2O_3$ phase. The $Fe_3O_4$ (111) films grown from thermal reduction show high crystallinity, even though films contain significant grain boundaries due to the larger mismatch between the in-plane unit cells of $Al_2O_3$ (001) and $Fe_3O_4$ (111).

## Acknowledgement

This project was primarily supported by the National Science Foundation (NSF), DMR under Award DMR-1454618. Additional support (X. Z.) was from NSF, DMR under Award DMREF: SusChEM 1436385. Z.M.Y and S.Y. acknowledge the support from the National Science Foundation of China (NSFC No. 51272209, 51471125 and 51501140), the Shaanxi Province Science and Technology Innovation Team Project (2013KCT-05). X.Z. is grateful for the help from Jack Rodenburg and Shi Cao on the MOKE measurements.

## Tables

**Table I.** The structures of the substrates and epitaxial orientations with different material during the deposition: $Al_2O_3$ (001), $\alpha$-$Fe_2O_3$ (001), $\gamma$-$Fe_2O_3$ (111) and $Fe_3O_4$ (111).[7,31–33]

|  | Structure | Lattice constants (bulk, in Å) | Lattice constant (plane /in Å) | Epitaxial orientation |
|---|---|---|---|---|
| $Al_2O_3$ | R-3c (167) | a=4.7602, c=12.9933 | (001), 4.7602 | (001), <100> / <120> |
| $\alpha$-$Fe_2O_3$ | R-3c (167) | a=5.007, c=13.641 | (001), 5.007 | (001), <100> / <120> |
| $\gamma$-$Fe_2O_3$ | $P4_332$ (212) | a=8.33 | (111), 11.78 | (111), <-211> / <01-1> |
| $Fe_3O_4$ | Fd-3m (227) | a=8.378 | (111), 11.85 | (111), <-211> / <01-1> |

**Table II.** Thermodynamic data used to calculate the Gibbs free energy change of the reaction from $\alpha$-$Fe_2O_3$ to $Fe_3O_4$.[37]

| $\Delta_fG^0$ (T) J/(mol K) | $\Delta_fG^0 = \Delta_fH^0 - T\Delta_fS^0$ |
|---|---|
| $\alpha$-$Fe_2O_3$ | -824640 – T 87.4 |
| $Fe_3O_4$ | -1115726 – T 146.14 |
| $\Delta_rG^0$ (T) | 80823 - 44.2 T |

**Table III**. Thermodynamic data used to calculate the Gibbs free energy change.[37]

|  | $\Delta_f H^0$ (kJ/mol) | $\Delta_f S^0$ (J/molK) |
|---|---|---|
| α-$Fe_2O_3$ | -824.640 | 87.4 |
| $Fe_3O_4$ | -1115.726 | 146.14 |
| $O_2$ | 0.0 | 205.15 |

**Table IV**. The parameters found in fitting the time constants of the α-$Fe_2O_3$ (001)→$Fe_3O_4$ (111) transition using Arrhenius law

| Pressure (Torr) | $\tau_0$ (sec) | Activation energy (eV) |
|---|---|---|
| $7.2 \times 10^{-6}$ | $7.3 \pm 7.1 \times 10^{-7}$ | $2.0 \pm 0.4$ |
| $6.4 \times 10^{-7}$ | $4.7 \pm 0.3 \times 10^{-9}$ | $2.5 \pm 0.6$ |
| $9.2 \times 10^{-8}$ | $8.9 \pm 5.7 \times 10^{-9}$ | $2.6 \pm 0.6$ |

Figures and captions

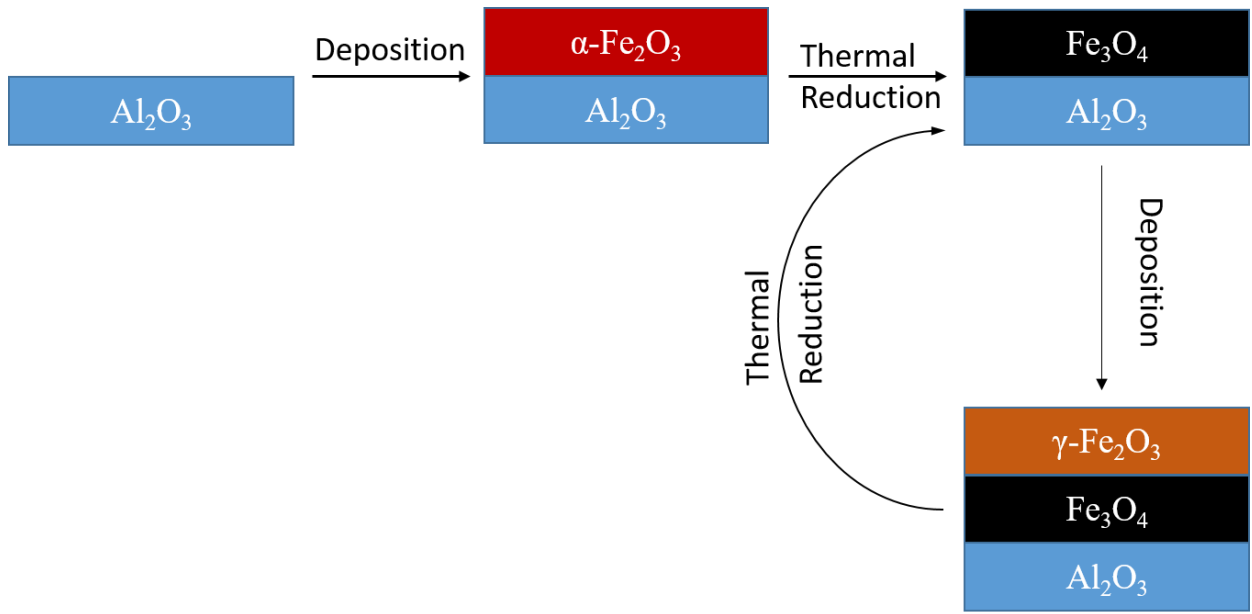

FIG 1 (color online). Schematic illustration of the deposition/thermal reduction processes to grow the $Fe_3O_4$ film.

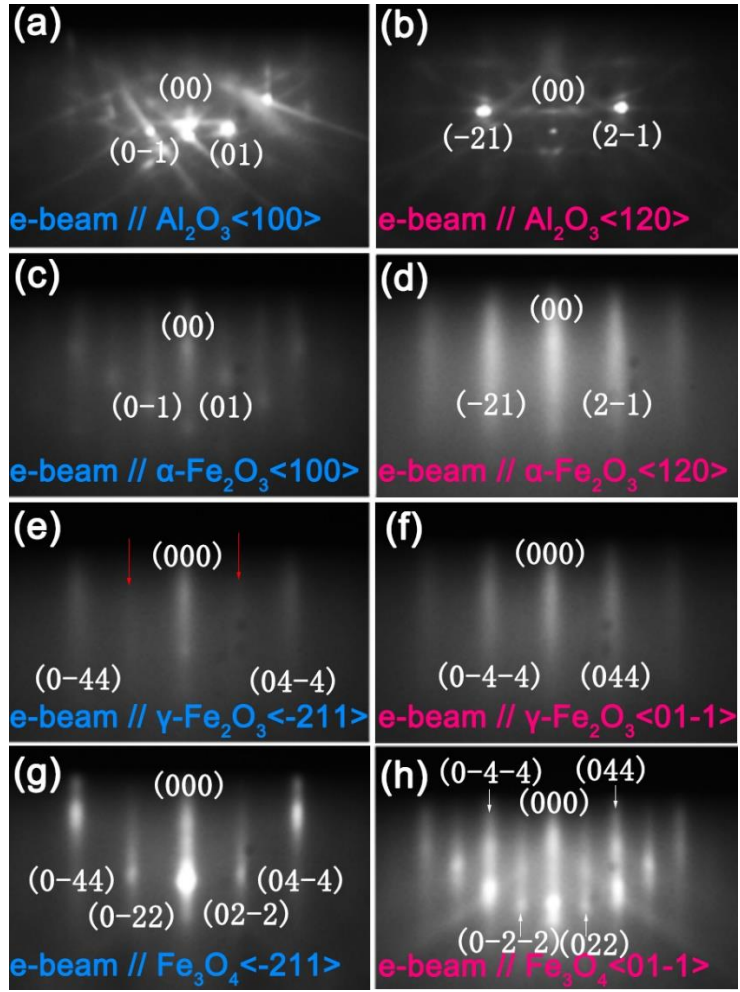

FIG 2 (color online). RHEED images of different surfaces with two different directions of incident electron beam. The diffraction streaks are marked using their corresponding reciprocal indices. (a) and (b) are the RHEED patterns of the $Al_2O_3$ (001) surface with e-beam parallel to $Al_2O_3$ <100> and <120>. (c) and (d) are the RHEED patterns of the $\alpha$-$Fe_2O_3$ (001) with e-beam parallel to $\alpha$-$Fe_2O_3$ <100> and <120>. (e) and (f) are the RHEED patterns of the $\gamma$-$Fe_2O_3$ (111) surface with e-beam parallel to $\gamma$-$Fe_2O_3$ <-211> and <01-1>. (g) and (h) are the RHEED patterns of the $Fe_3O_4$ (111) surface with e-beam parallel to $Fe_3O_4$ <-211> and <01-1>. The direction of the electron beam are the same for the images (a), (c), (e), and (g), and the same for the images (b), (d), (f), and (h).

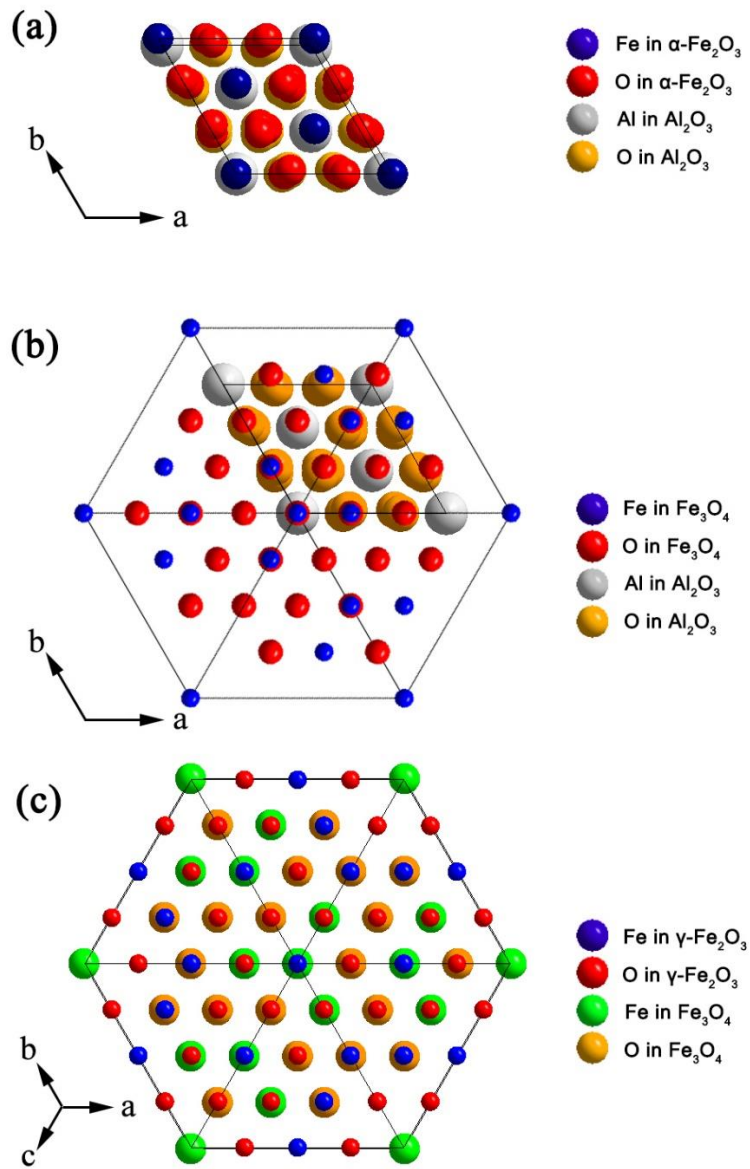

FIG 3 (color online). Schematics of the epitaxial relations. (a) Between $Al_2O_3$ (001) and $\alpha$-$Fe_2O_3$ (001). (b) Between $Al_2O_3$ (001) and $Fe_3O_4$ (111). (c) Between $\gamma$-$Fe_2O_3$ (001) and $Fe_3O_4$ (111).

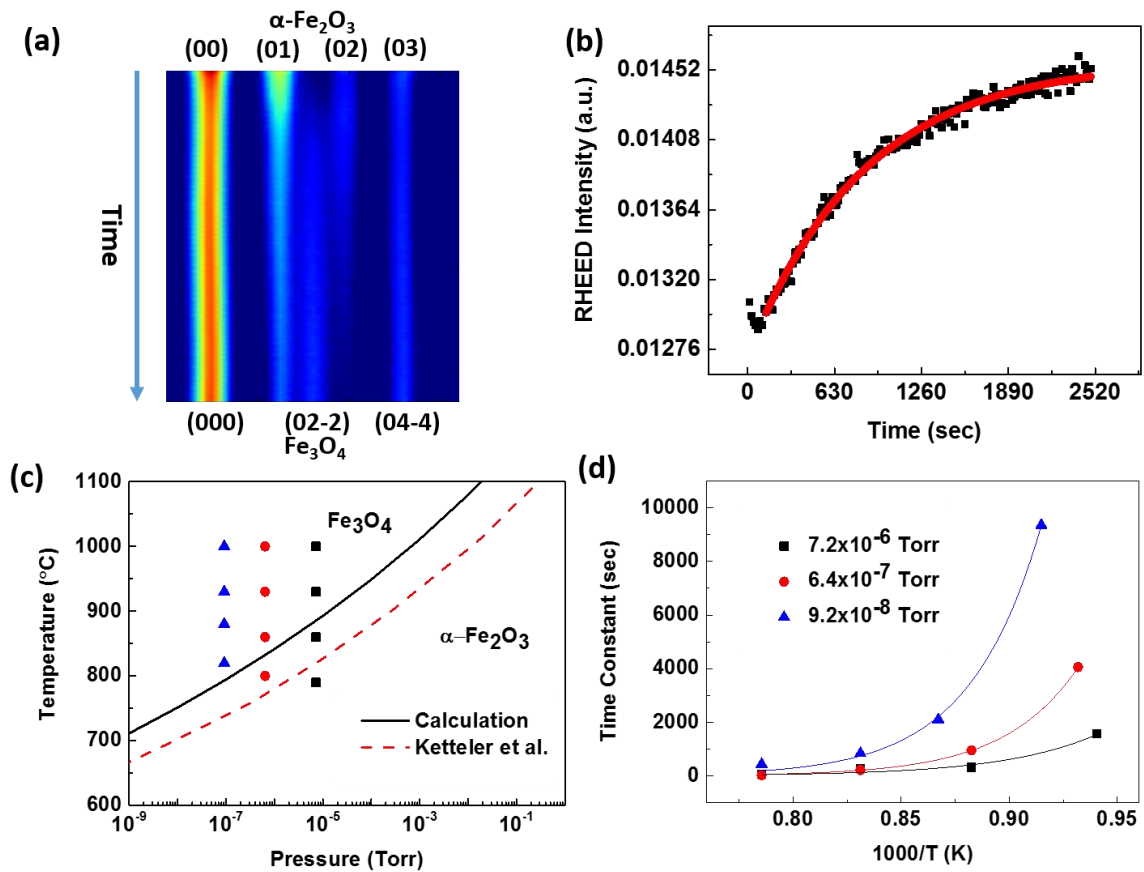

FIG 4 (color online). Thermal dynamics and kinetics of the α-$Fe_2O_3$ (001)→$Fe_3O_4$ (111) transition on the $Al_2O_3$ (001) substrates. (a) Time evolution of the RHEED pattern (see text) at 930 °C in 9.2×$10^{-8}$ Torr $O_2$. (b) The intensity of the $Fe_3O_4$ (02-2) streak as a function of time calculated from (a) and the fit (line). (c) The thermodynamic calculation of the phase boundary between α-$Fe_2O_3$ and $Fe_3O_4$ and the conditions for the experimental measurements. The dashed line is the calculation from Ketteler et al.[38] (d) The temperature dependence of time constant of the α-$Fe_2O_3$ (001) → $Fe_3O_4$ (111) transition and the fit (lines) using the Arrhenius law.

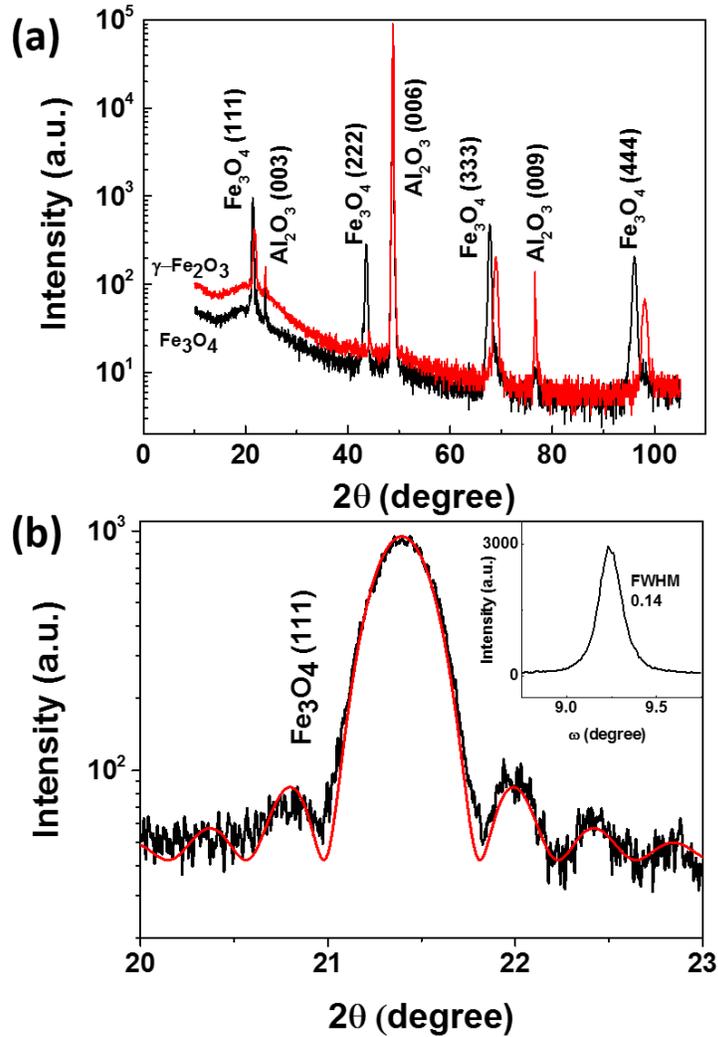

FIG 5 (color online). X-ray diffraction of the $Fe_3O_4$ film (~30 nm) as well as the annealed $Fe_3O_4$ film ($\gamma$-$Fe_2O_3$, see text). (a) Large range $\theta$-$2\theta$ scan using a cobalt K-$\alpha$ source ($\lambda$=1.79 Å). The indices of the $\gamma$-$Fe_2O_3$ peaks are the same as those of nearest $Fe_3O_4$ peaks. (b) The close-up view of the $Fe_3O_4$ (111) diffraction peak. The line is the fit of the Laue oscillation (see text). The inset is the rocking curve of the $Fe_3O_4$ (111) peak measured with a Cu K-$\alpha$ source ($\lambda$=1.54 Å). The full-width-half-maximum (FWHM) of the rocking curve is about 0.14 degree.

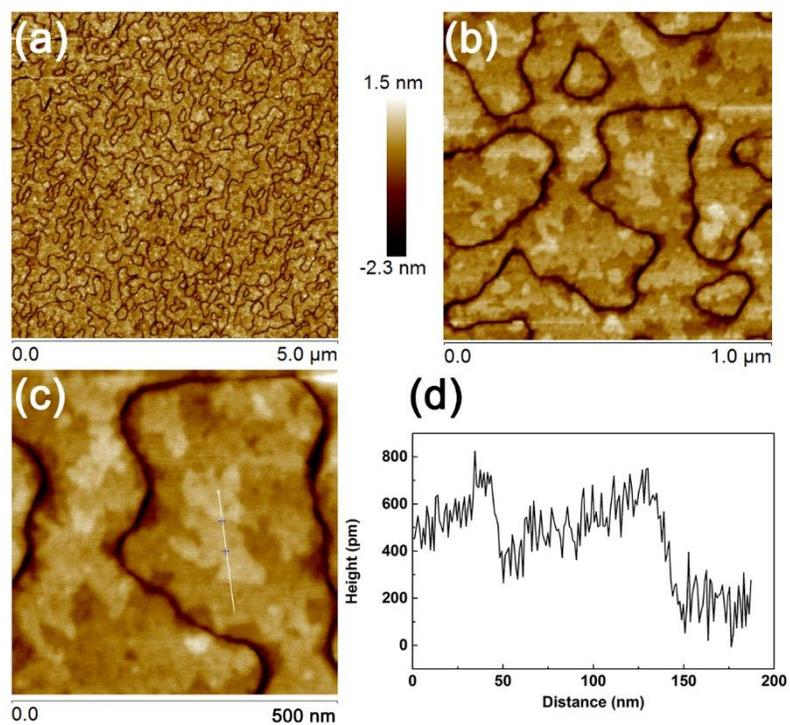

FIG 6 (color online). Surface morphology of the $Fe_3O_4$ film (~30 nm) measured using atomic force microscopy. (a) 5×5 μm. (b) 1×1 μm and (c) 500×500 nm. (d) A cross section of the film surface indicated by the line in (c).

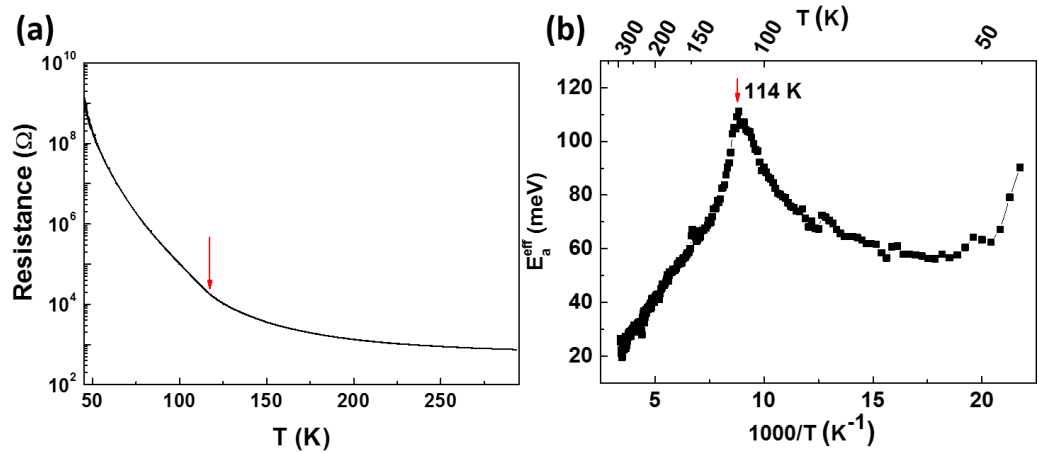

FIG 7 (color online). Temperature dependent electrical resistance (a) and the effective activation energy (b) (see text) of the $Fe_3O_4$ film (~30 nm). The Verwey transition is marked using the vertical arrows.

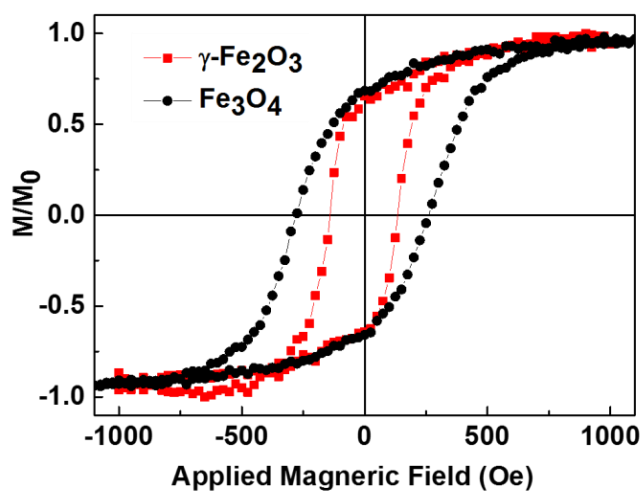

FIG 8 (color online). Magneto optical Kerr effect of the $Fe_3O_4$ film (~30 nm) and the $\gamma$-$Fe_2O_3$ film (from annealing a $Fe_3O_4$ film in $O_2$, see text), measured at room temperature.

# Supplementary materials

Kinetics and Intermediate Phases in Epitaxial Growth of $Fe_3O_4$ Films from Deposition and Thermal Reduction

## Surface morphology of the $Fe_3O_4$ (111) films measured using atomic force microscopy (AFM)

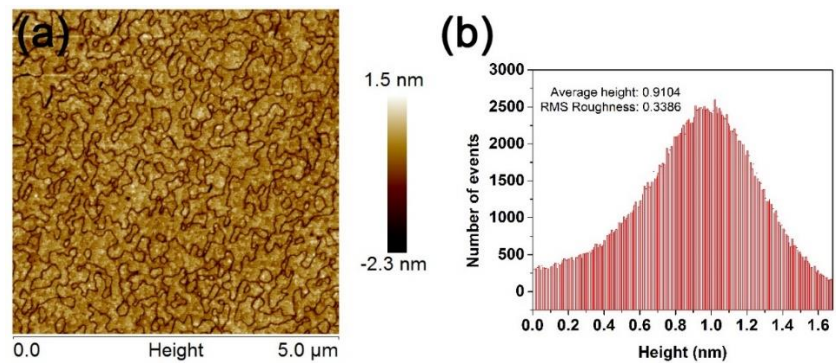

**Figure S1.** (a) AFM image of an $Fe_3O_4$ (111) film. (b) Height distribution and room-mean-square (RMS) roughness of the $Fe_3O_4$ (111) film.

# Transition of the surface structure from the time evolution of the reflection high energy electron diffraction (RHEED) patterns

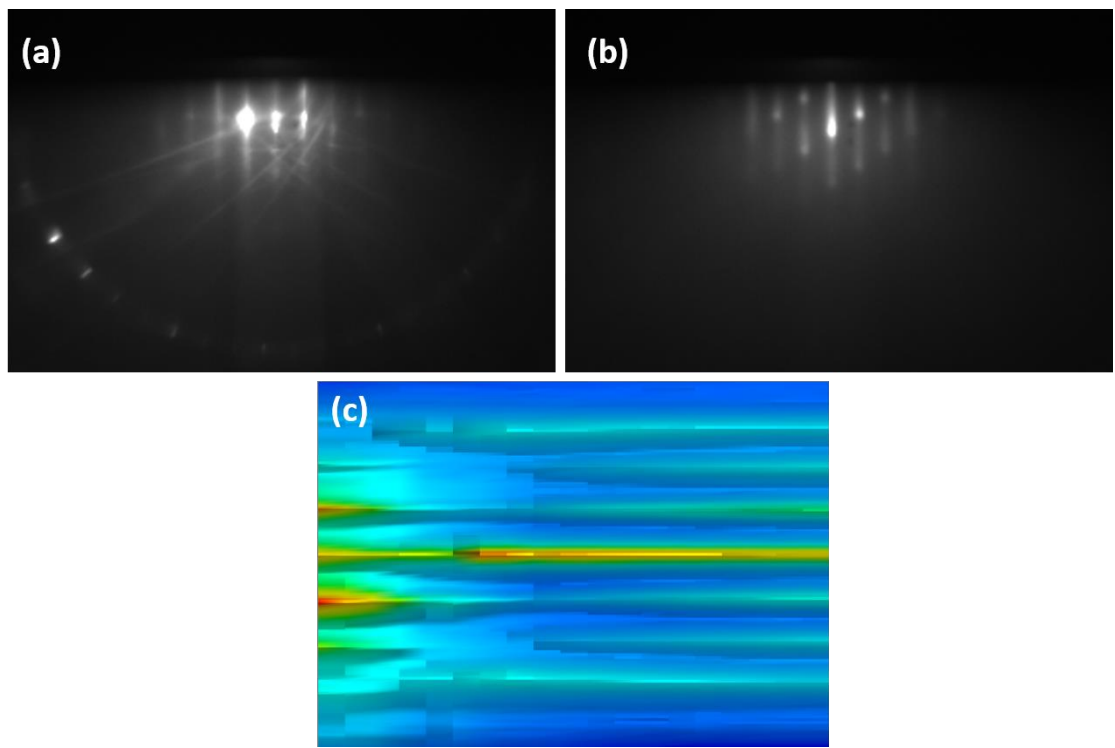

**Figure S2.** (a) RHEED pattern of the Al$_2$O$_3$ (001) substrate. (b) The RHEED pattern of α-Fe$_2$O$_3$ (001) deposited on the Al$_2$O$_3$ substrate. (c) The RHEED evolution during the deposition. The horizontal axis is the time.

Figure S2 shows the time evolution of the RHEED pattern during the deposition of an α-Fe$_2$O$_3$ layer on the Al$_2$O$_3$ (0001) substrate. The deposition started with a bare Al$_2$O$_3$ substrate, the RHEED pattern of which is shown in the Fig. S2 (a) with the incident e-beam along Al$_2$O$_3$ <100>. The same diffraction pattern is shown in the Fig. 2(a) in the main text. After the deposition of a 2.5 nm Fe$_2$O$_3$ layer at 300 °C, the RHEED pattern becomes the one shown in Fig. S2(b), which is similar but with a different lattice constant, indicating a α-Fe$_2$O$_3$ (001) layer. The evolution of the RHEED pattern during the deposition is shown in Fig. S2(c).

The evolution of the RHEED pattern (Fig. S2(c)) is obtained using the following procedure. 1) RHEED images are taken every 15 seconds. 2) A region of interest (ROI) containing the first order of the diffraction streaks is chosen (see Fig. S3). 3) The intensity of the ROI is integrated along the longer direction of the streaks to get a diffraction spectrum. 4) The diffraction spectra are plotted again time into a two dimensional image, to show the time evolution of the diffraction.

A clear transition of the surface from Al$_2$O$_3$ (0001) to Fe$_2$O$_3$ (001) can be visualize in the time evolution of the RHEED patterns. The time scale for the transition can also be analyzed.

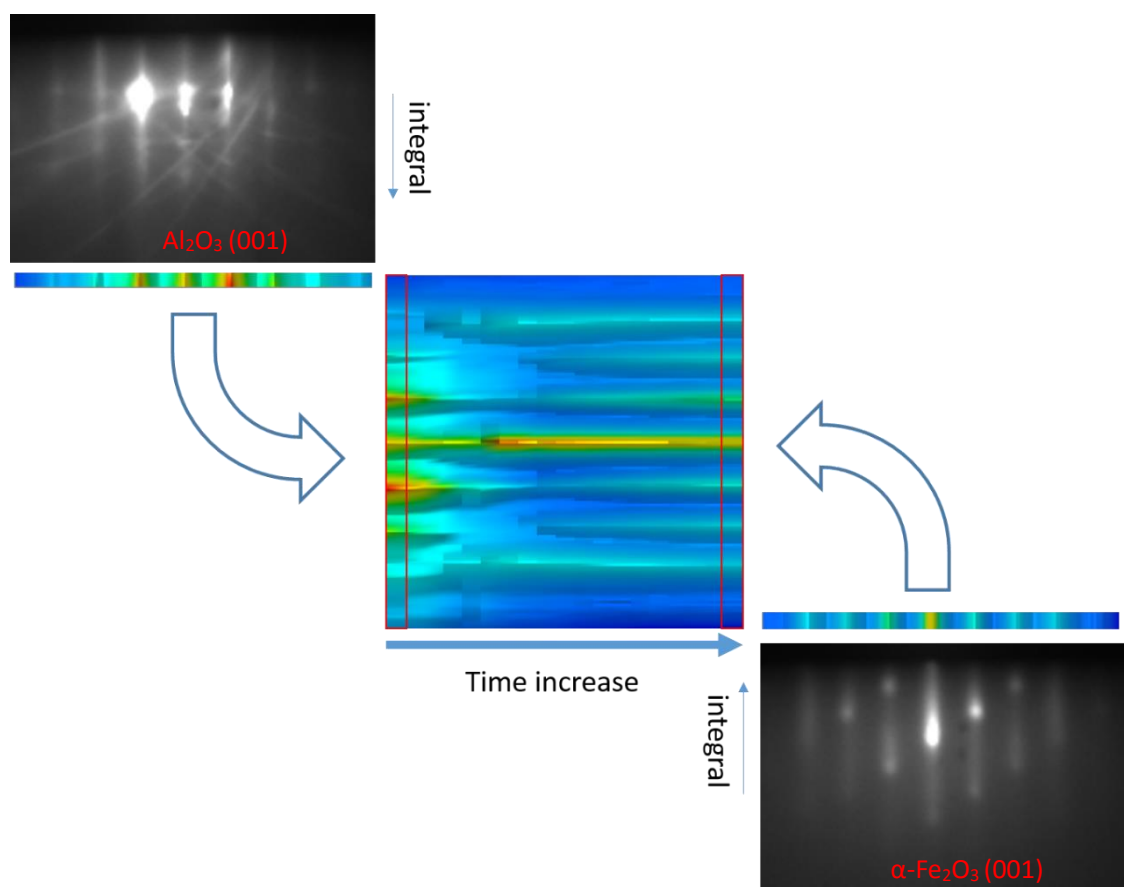

**Figure S3.** Schematic illustration of how to obtain the time evolution of the RHEED patterns

# Comparison of the γ-Fe₂O₃ and Fe₃O₄ phases using the X-ray diffraction (XRD) measurements

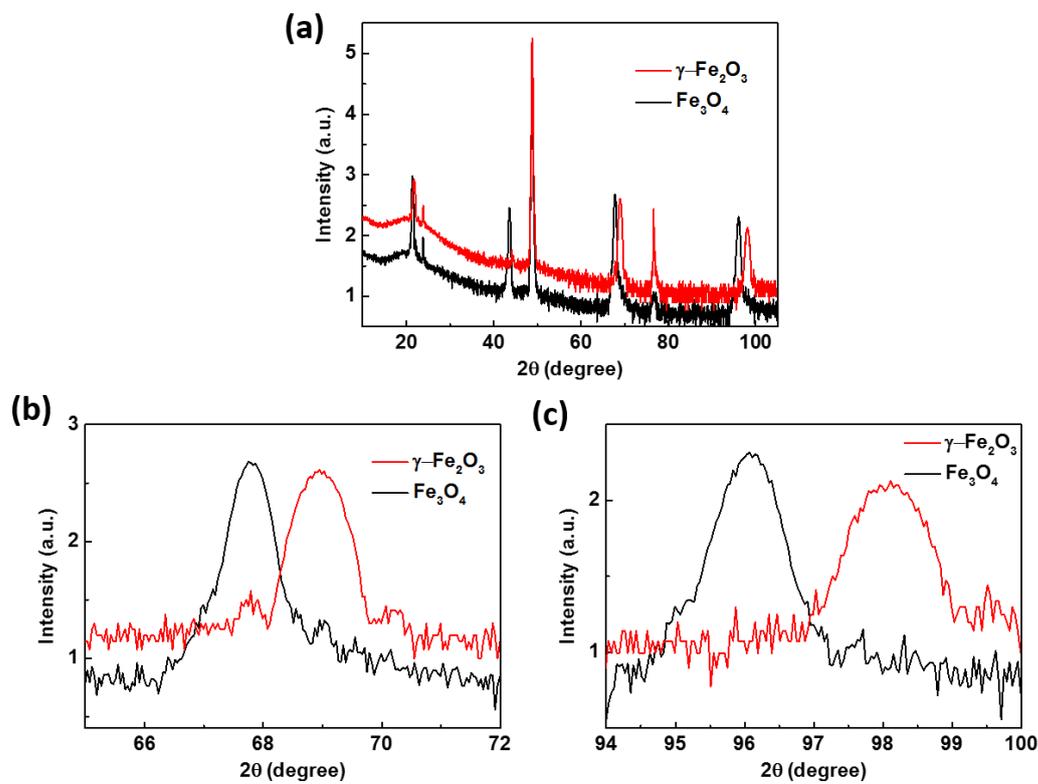

**Figure S4.** (a) The XRD spectra of Fe$_3$O$_4$ (111) and γ-Fe$_2$O$_3$ (111) films. (b) The close-up view of the spectra near the (333) (b) peak and the (444) peak (c) of Fe$_3$O$_4$ (111) and γ-Fe$_2$O$_3$ (111) films.

The similarity of the XRD spectra of the two films indicate similar crystal structures, which are the inverse spinel structure (space group = Fd-3m) and the cation deficient spinel structure for Fe$_3$O$_4$ and γ-Fe$_2$O$_3$ respectively. The lattice parameters are 8.378 Å and 8.33 Å for Fe$_3$O$_4$ and γ-Fe$_2$O$_3$ respectively. Because γ-Fe$_2$O$_3$ has a smaller lattice constant, the Bragg angles of the γ-Fe$_2$O$_3$ diffraction are larger than the corresponding angles of the Fe$_3$O$_4$ diffraction. According to Fig. S4(b) and S4(c), we can see the differences of the Bragg angles are 1.2 degree and 2.04 degree for the (333) and (444) peaks respectively.

# $Fe_3O_4$ and $\gamma$-$Fe_2O_3$ d-spacing calculated from the X-ray diffraction measurements

**Table S(I).** $Fe_3O_4$ lattice parameters calculated from x-ray diffraction.

| Peak/Plane | 2θ (deg) | d (Å) | $d_{(111)}$ (Å) |
|---|---|---|---|
| (111) | 21.432 | 4.811 | 4.811 |
| (222) | 43.592 | 2.409 | 4.818 |
| (333) | 67.752 | 1.605 | 4.814 |
| (444) | 96.072 | 1.203 | 4.812 |

**Table S(II).** $\gamma$-$Fe_2O_3$ lattice parameters calculated from x-ray diffraction.

| Peak/Plane | 2θ (deg) | d (Å) | $d_{(111)}$ (Å) |
|---|---|---|---|
| (111) | 21.712 | 4.749 | 4.749 |
| (222) | 44.152 | 2.380 | 4.760 |
| (333) | 68.952 | 1.580 | 4.741 |
| (444) | 98.112 | 1.184 | 4.737 |

# RHEED patterns of the γ-Fe₂O₃ (111) film obtained by annealing a Fe₃O₄ (111) film

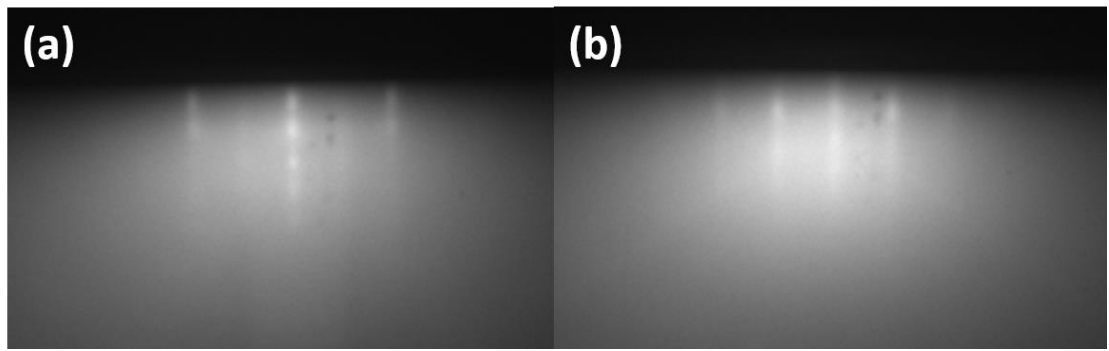

**Figure S5.** The RHEED pattern of a γ-Fe₂O₃ (111) film (from annealing a Fe₃O₄ (111) film) with the electron beam along <-211> (a) and <01-1> (b) direction.

The RHEED pattern of the annealed film is similar to the RHEED pattern shown in Fig. 2(e) and 2(f) in the main text. These results indicate that the intermediate product during the deposition onto the Fe₃O₄ (111) surface is γ-Fe₂O₃ (111).

## Derivation of Gibbs free energy change

Definition of symbols:
$G$: Gibbs free energy
$\Delta_f G$: formation Gibbs free energy. This is the Gibbs free energy of a compound relative to the corresponding elemental matters.
$\Delta_f G^0$: formation Gibbs free energy at standard condition ($P=1.0\times10^5$ Pa)
$\Delta_r G$: Gibbs free energy change for a reaction
$\Delta_r G^0$: Gibbs free energy change for a reaction at the standard condition

Consider the reaction: $\alpha - Fe_2O_3 \rightleftharpoons \frac{2}{3} Fe_3O_4 + \frac{1}{6} O_2$,

the Gibbs free energy change is $\Delta_r G = (\frac{2}{3} G_{Fe3O4} + \frac{1}{6} G_{O2}) - G_{Fe2O3}$.

Since $G_{Fe2O3}$, $G_{Fe3O4}$, and $G_{O2}$ are not generally available, we will try to derive their relation to the values at the standard condition.
From

$$G_{Fe2O3} = \Delta_f G_{Fe2O3} + 2G_{Fe} + \frac{3}{2} G_{O2}$$

$$G_{Fe3O4} = \Delta_f G_{Fe3O4} + 3G_{Fe} + 2G_{O2},$$

one finds

$$\Delta_r G = \left[\frac{2}{3}(\Delta_f G_{Fe3O4} + 3G_{Fe} + 2G_{O2}) + \frac{1}{6} G_{O2}\right] - (\Delta_f G_{Fe2O3} + 2G_{Fe} + \frac{3}{2} G_{O2})$$

$$= \frac{2}{3} \Delta_f G_{Fe3O4} - \Delta_f G_{Fe2O3}.$$

Now we look at,
$$\Delta_f G_{Fe3O4} = \Delta_f G^0_{Fe3O4} + (G_{Fe3O4} - G^0_{Fe3O4}) - 3(G_{Fe} - G^0_{Fe}) - 2(G_{O2} - G^0_{O2}).$$
Since for solid state matters, the Gibbs free energy is not affected by the pressure very much, one can assume $G_{Fe3O4} - G^0_{Fe3O4} = 0$ and $G_{Fe} - G^0_{Fe} = 0$. Therefore, it follows from that
$$\Delta_f G_{Fe3O4} = \Delta_f G^0_{Fe3O4} - 2(G_{O2} - G^0_{O2}).$$

Similarly, one has $\Delta_f G_{Fe2O3} = \Delta_f G^0_{Fe2O3} - \frac{3}{2}(G_{O2} - G^0_{O2})$.

Now the Gibbs free energy change of the reaction becomes
$$\Delta_r G = \frac{2}{3}[\Delta_f G^0_{Fe3O4} - 2(G_{O2} - G^0_{O2})] - [\Delta_f G^0_{Fe2O3} - \frac{3}{2}(G_{O2} - G^0_{O2})] = \frac{2}{3} \Delta_f G^0_{Fe3O4} + \frac{1}{6}(G_{O2} - G^0_{O2}) - \Delta_f G^0_{Fe2O3}.$$

If the $O_2$ is treated as an ideal gas, then the molar Gibbs free energy change is $G_{O2} - G^0_{O2} = RT \ln\left(\frac{P}{P_0}\right)$, where $P_0 = 1.0\times10^5$ Pa is the pressure at standard condition.

Hence, one reaches the relation: $\Delta_r G = \frac{2}{3} \Delta_f G^0_{Fe3O4} - \Delta_f G^0_{Fe2O3} + \frac{1}{6} RT \ln\left(\frac{P}{P_0}\right)$.

## Calculation of Gibbs free energy change at standard condition

**Table S(III).** Data for calculating the Gibbs free energy change.[1]

|  | $\Delta_f H^0$ (kJ/mol) | $\Delta S^0$ (J/mol K) |
|---|---|---|
| α-$Fe_2O_3$ | -824.640 | 87.4 |
| $Fe_3O_4$ | -1115.726 | 146.14 |
| $O_2$ | 0.0 | 205.15 |

$\Delta_r G^0(T) = \Delta_r H^0 - T\Delta_r S^0$

$= \left(\frac{2}{3}\Delta_f H^0_{Fe3O4} + \frac{1}{6}\Delta_f H^0_{O2} - \Delta_f H^0_{Fe2O3}\right)$

$\qquad - T\left(\frac{2}{3}\Delta_f S^0_{Fe3O4} + \frac{1}{6}\Delta_f S^0_{O2} - \Delta_f S^0_{Fe2O3}\right)$

$= \left(\frac{2}{3} \times (-1115726) + \frac{1}{6} \times 0 + 824640\right)$

$\qquad - T\left(\frac{2}{3} \times 146.14 + \frac{1}{6} \times 205.15 - 87.4\right)$

$= (80823 - 44.2\,T)$ J/mol

**The crystallite size in the out-of-plane direction of a Fe₃O₄ (111) film**

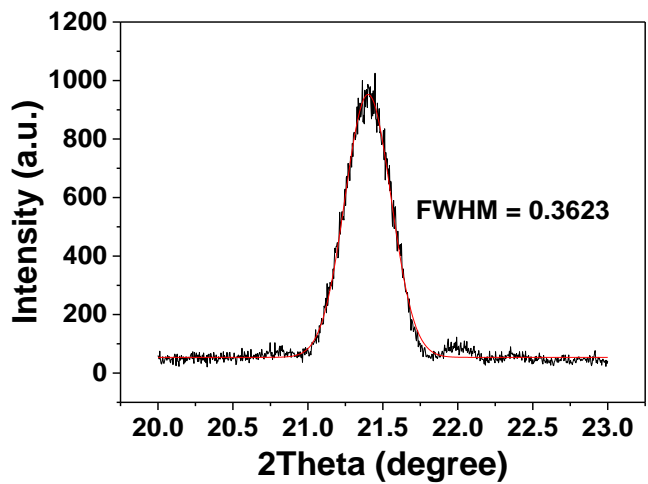

**Figure S6.** The (111) peak of the Fe₃O₄ thin film.

The size of the crystallites can be estimated using the Scherrer equation[2] $\tau = \frac{K\lambda}{\beta cos(\theta)}$, where $\tau$ is the mean size of the crystalline domains, $K$ is a dimensionless shape factor (taken as 0.9), $\lambda$ is the X-ray wavelength, $\beta$ is width of the diffraction peak (FWHM) in radians, and $\theta$ is the Bragg angle.

According to Fig. S6, one finds

$$\tau = \frac{K\lambda}{\beta cos(\theta)} = \frac{0.9 * 1.79}{0.3623 * pi/180 * cos(21.4015/2 * pi/180)} = 259.3 \text{ Å}.$$

# The crystallite size of the in-plane direction of Fe₃O₄ (111)

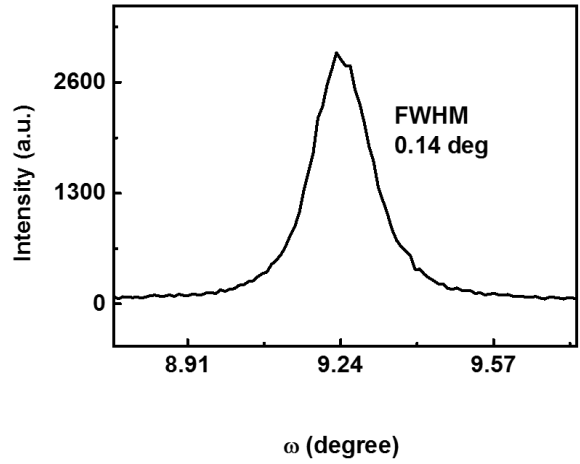

**Figure S7.** The rocking curve of the Fe₃O₄ thin film.

Using Scherrer equation $\tau = \frac{K\lambda}{\beta \cos(\theta)}$, one finds,

$$\tau = \frac{K\lambda}{\beta \cos(\theta)} = \frac{0.9 * 1.54}{0.14 * pi/180 * \cos(9.2309 * pi/180)} = 574.7 \text{ Å}.$$